\documentclass[preprint,12pt]{elsarticle}
\usepackage{bbold}
\usepackage{graphicx}
\usepackage{amssymb}
\usepackage{amsmath}
\usepackage{empheq}
\usepackage{amsthm}
\usepackage{siunitx}

\usepackage{dcolumn}
\usepackage{bm}
\newcommand{\bfkappa}{\boldsymbol{\kappa}}
\usepackage{color}
\usepackage{float}

\begin{document}


\title{Topological rainbow trapping for elastic energy harvesting in graded SSH systems}

\author[label1]{Gregory~J.~Chaplain}  
\author[label2,label3]{Jacopo M. De Ponti}
\author[label4]{Giulia Aguzzi}
\author[label4]{Andrea Colombi}
\author[label1,label5,label6]{Richard~V.~Craster}


\address[label1]{Department of Mathematics, Imperial College London, 180 Queen's Gate, South Kensington, London SW7 2AZ}
\address[label2]{Dept. of Civil and Environmental Engineering, Politecnico di Milano, Piazza Leonardo da Vinci, 32, 20133 Milano, Italy}
\address[label3]{Dept. of Mechanical Engineering, Politecnico di Milano, Via Giuseppe La Masa, 1, 20156 Milano, Italy}
\address[label4]{Dept. of Civil, Environmental and Geomatic Engineering, ETH, Stefano-Franscini-Platz 5, 8093 Z\"urich, Switzerland}
\address[label5]{Department of Mechanical Engineering, Imperial College London, London SW7 2AZ, UK}
\address[label6]{UMI 2004 Abraham de Moivre-CNRS, Imperial College London, London SW7 2AZ, UK}

\begin{abstract}
We amalgamate two fundamental designs from distinct areas of wave control in physics, and place them in the setting of elasticity. Graded elastic metasurfaces, so-called metawedges, are combined with the now classical Su-Schrieffer-Heeger (SSH) model from the field of topological insulators. The resulting structures form one-dimensional graded-SSH-metawedges that support multiple, simultaneous, topologically protected edge states. These robust, enhanced localised modes are leveraged for applications in elastic energy harvesting using the piezoelectric effect. The designs we develop are first motivated by applying the SSH model to mass-loaded Kirchhoff-Love thin elastic plates. We then extend these ideas to using graded resonant rods, and create SSH models, coupled to elastic beams and full elastic half-spaces. 
\end{abstract}

\maketitle

\section{Introduction}

Topological insulators are exotic materials in which protected edge or interfacial surface states exist between bulk band gaps, owing their existence to broken symmetries within a periodic system. Despite their origins in quantum mechanical systems \cite{moore2010birth, hasan2010colloquium,qi2011topological}, there has been a flurry of intensive research translating these effects into all flavours of classical wave propagation, from electromagnetism and acoustics to mechanics and elasticity \cite{yang2015topological,chen2018study}; the protected edge modes can have attractive properties such as resilience to backscattering from defects and impurities, and can exhibit unidirectional propagation. As such the physical phenomena surrounding topological insulators has naturally led to a concerted effort mapping such effects into metamaterial and photonic crystal (and their analogues) design \cite{khanikaev2013photonic, mousavi2015topologically, susstrunk2015observation}. 

The nature of the symmetry breaking giving rise to a protected edge state defines two classes of topological insulators. Inspired by the original quantum mechanical systems exhibiting the quantum Hall effect (QHE), where time reversal symmetry (TRS) is broken through applied external fields \cite{klitzing1980new, haldane1988model}, has led to so-called active topological materials \cite{wang2015topological,NASH2015, souslov2017topological, zhang2018topological, souslov2019topological}. Similarly the quantum spin Hall effect (QSHE), in which symmetry breaking is achieved through spin-orbit interactions (TRS is preserved)\cite{kane2005quantum,bernevig2006quantum}, has engendered passive topological systems. Such systems have promulgated simpler topological insulator motivated designs in continuum wave systems through the breaking of geometric symmetries to induce topologically nontrivial bandgaps  \cite{makwana2018geometrically}. 

Underpinning the topological nature of the Bloch bands defined by each material are associated invariants which characterise the geometric phase, that is the phase change associated with a continuous, adiabatic deformation of the system; most notably the Berry phase \cite{berry1984quantal, xiao2010berry}, and its one dimensional counterpart, the Zak phase \cite{Zak1989}.

Numerous translations of 2D topological insulators to wave physics have been realised, often based around honeycomb structures \cite{wu2015scheme}, which guarantee symmetry-induced Dirac points that can be leveraged to induce edge states at the interface between two topologically distinct media. These have been replicated for waveguiding applications for photonics \cite{khanikaev2017two}, phononics \cite{lu2016valley} and platonics \cite{pal2017edge, chaunsali2018subwavelength}. More nuanced interpretations have achieved beam splitter designs with square lattices \cite{Makwana2019tuneable,makwana2019experimental,makwana2019topological}. Higher order topological effects, for higher dimensional structures have also received much attention \cite{ni2019observation,li2020higher}.

Despite their relative simplicity, 1D topological insulators serve not only as pedagogical examples, but posses important features for applications ranging from lasing \cite{parto2018edge} to mechanical transport \cite{chen2014nonlinear}. Motivated by applications in elastic energy harvesting, we highlight a new modality of 1D topological insulators, based on the well-established Su-Schrieffer-Heeger (SSH) model \cite{su1979solitons}, via the amalgamation of this model with graded metawedge structures \cite{colombi2016seismic,Colombi2017,skelton2018multi,chaplain2019tailored}. 

\begin{figure}[t]
    \centering
    \includegraphics[width = 0.8\textwidth]{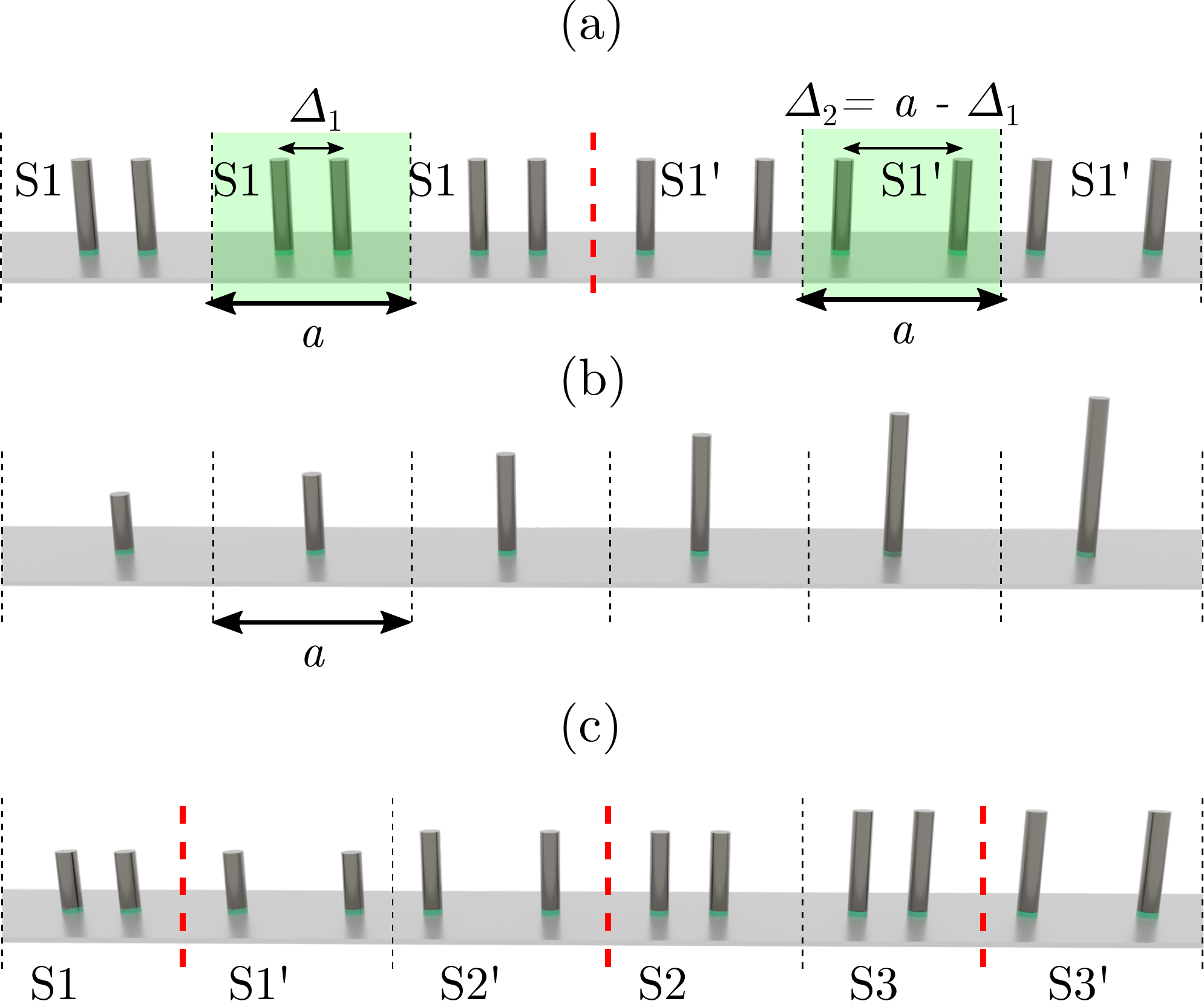}
    \caption{Combining SSH interfaces with metawedge structures: (a) Shows an elastic version of the SSH model, where an interface is encountered between structures S1 and S1$^\prime$. The unit cells of each are highlighted in green, each shares the same periodicity $a$, but have rod separation $\Delta_{1}$ and $\Delta_{2}$ respectively. (b) Shows conventional metawedge structure  consisting of periodically spaced rods of increasing height. (c) Shows the amalgamation of these geometries to produce several SSH interfaces for differing rods heights, marked by the dashed red lines. The green material at the base of each rod represents a piezoelectric material which is to be used for energy harvesting \cite{DePonti2019,chaplain2020rainbow}.}
    \label{fig:IntroFig}
\end{figure}

Throughout this article we will examine the SSH model, due to its successful predictions and ease of its translation from topological wave physics to mechanical systems \cite{huber2016topological}, but now in the setting of elasticity. Figure~\ref{fig:IntroFig} highlights the overarching theme of this article, which combines the conventional SSH model with a graded system. Shown in Figure~\ref{fig:IntroFig}(a) is an elastic version of the classical SSH interface, for resonant rods atop a beam. The interface at which a topological edge mode exists is highlighted by the red dashed line; this is where two related geometries meet. To the left of this interface are a set of periodic unit cells, of width $a$, each with two resonant rods set apart a distance $\Delta_{1}$ from the centre of the cell: we call this structure S1. To the right of the interface we consider structures, S1$^\prime$, that consist of unit cells of the same width, but this time with the rods placed a distance $\Delta_{2} = a-\Delta_{1}$ apart from the cell centre. In an infinite periodic array both these structures are identical as there is merely a translation in the definition of the unit cell. For that infinite array, then, taking advantage of periodicity to introduce Floquet-Bloch waves, both structures have the same dispersion curves. We show in Section~\ref{sec:SSH_KL} how an edge mode arises at this interface, sometimes referred to as a domain wall.

Shown in Figure~\ref{fig:IntroFig}(b) is the, now conventional, graded metawedge structure \cite{colombi2016seismic}, that has recently been utilised for energy harvesting purposes \cite{DePonti2019,de2020experimental}. This consists of periodically spaced rods, which increase in height through an adiabatic grading. The utility of such devices is provided by their ability to manipulate and segregate frequency components by slowing down waves which can reach effective local bandgaps at different spatial positions \cite{chaplain2020rainbow}. Our  desire is to combine these two structures, as shown in Figure~\ref{fig:IntroFig}(c), to incorporate several, simultaneous, topologically protected edge modes for energy harvesting applications. Such a structure is devised by alternating between primed and un-primed pairs of structures for differing rod heights. 

To elucidate the design paradigm and conditions for existence of an edge mode, we firstly  consider the simplified elastic model of a point mass loaded thin Kirchhoff-Love elastic plate. The expected existence of edge states in a one dimensional SSH chain is confirmed through calculation of the Zak phase via an efficient numerical scheme \cite{chaplain2019rayleigh}, corroborated via Fourier spectral analysis of scattering simulations. The differences between the localised 1D edge states and that of conventional band gap defect states are highlighted. This methodology is then extended to a topological system of resonating rods atop an elastic beam. Recent experimental work has highlighted the existence of such states in quasi-periodic resonant-loaded beams \cite{xia2020topological}, whilst other works incorporate piezoelectric effects to tune the topological phases of the bands \cite{zhou2020actively}. Here we continue with the SSH model demonstrating that, by the addition of piezoelectric materials, efficient energy harvesting from mechanical to electric energy is possible; this extends the applications of coupling piezoelectricity with topological insulators \cite{mchugh2016topological}. The motivation of coupling with the graded structures, as highlighted in Figure~\ref{fig:IntroFig}, is to extend the bandwidth from the single frequency at which the edge mode exist thereby achieving  broadband performance of the device with an attractively compact device. Finally, this model is extended to elastic halfspaces that support Rayleigh waves, introducing the concept to broaden the scope of topological groundborne vibration control.

\section{SSH in Thin Elastic Plates}
\label{sec:SSH_KL}

The equations governing flexural wave propagation in thin Kirchhoff-Love (KL) elastic plates \cite{landau1986lifshitz} provide a flexible avenue for investigating a wide variety of wave manipulation effects; they efficiently predict wave behaviour in physical systems \cite{lefebvre2017unveiling}, with elegant solutions readily available for point loaded scatterers \cite{evans2007penetration}. Further to this, the Green's function of the governing biharmonic wave equation is non-singular and remains bounded, and as such numerical complications during the implementation of scattering simulations are side-stepped, enabling efficient scattering calculations to be obtained by extending a method attributed to Foldy \cite{foldy1945multiple}. Recent advances for analysing one-dimensional, infinite, periodic structures, in such systems, \cite{chaplain2019rayleigh} have generated efficient methods for calculating their dispersion curves, enabling fast design and analysis. These features of the KL system, and the numerical ease of its solution, motivate its use as a powerful toolbox for quickly characterising topological systems \cite{torrent13a,makwana2018geometrically,makwana2018designing,Makwana2019tuneable,chen2019mechanical,jin2020topological}.

For a point mass loaded KL plate, loaded with $J$ masses of value $M^{(j)}$ at positions $\boldsymbol{x}^{(j)}$, the restoring forces at the mass position are proportional to the displacement of the mass at that point, resulting in the  out-of-plane flexural wave displacement, $w(\boldsymbol{x})$, being governed by the biharmonic wave equation,
\begin{equation}
(\nabla^{4} - \Omega^{2})w(\boldsymbol{x}) = \Omega^2\sum\limits_{j = 1}^{J} M^{(j)}w(\boldsymbol{x})\delta(\boldsymbol{x}-\boldsymbol{x}^{(j)}).
\label{eq:KL}
\end{equation}
We adopt a non-dimensionalised frequency such that \(\Omega^2 = \rho h\omega^2/D\), where \(\rho\) is the mass density of the plate and \(h\) is the plate thickness, with \(\omega\) being the dimensional angular frequency. \(D\) is the flexural rigidity, which encodes the Young's modulus, \(E\), and Poisson's ratio, \(\nu\), of the plate through \(D = Eh^3/12(1-\nu^2)\). 

\begin{figure}[t!]
     \centering
     \includegraphics[width = 0.8\textwidth]{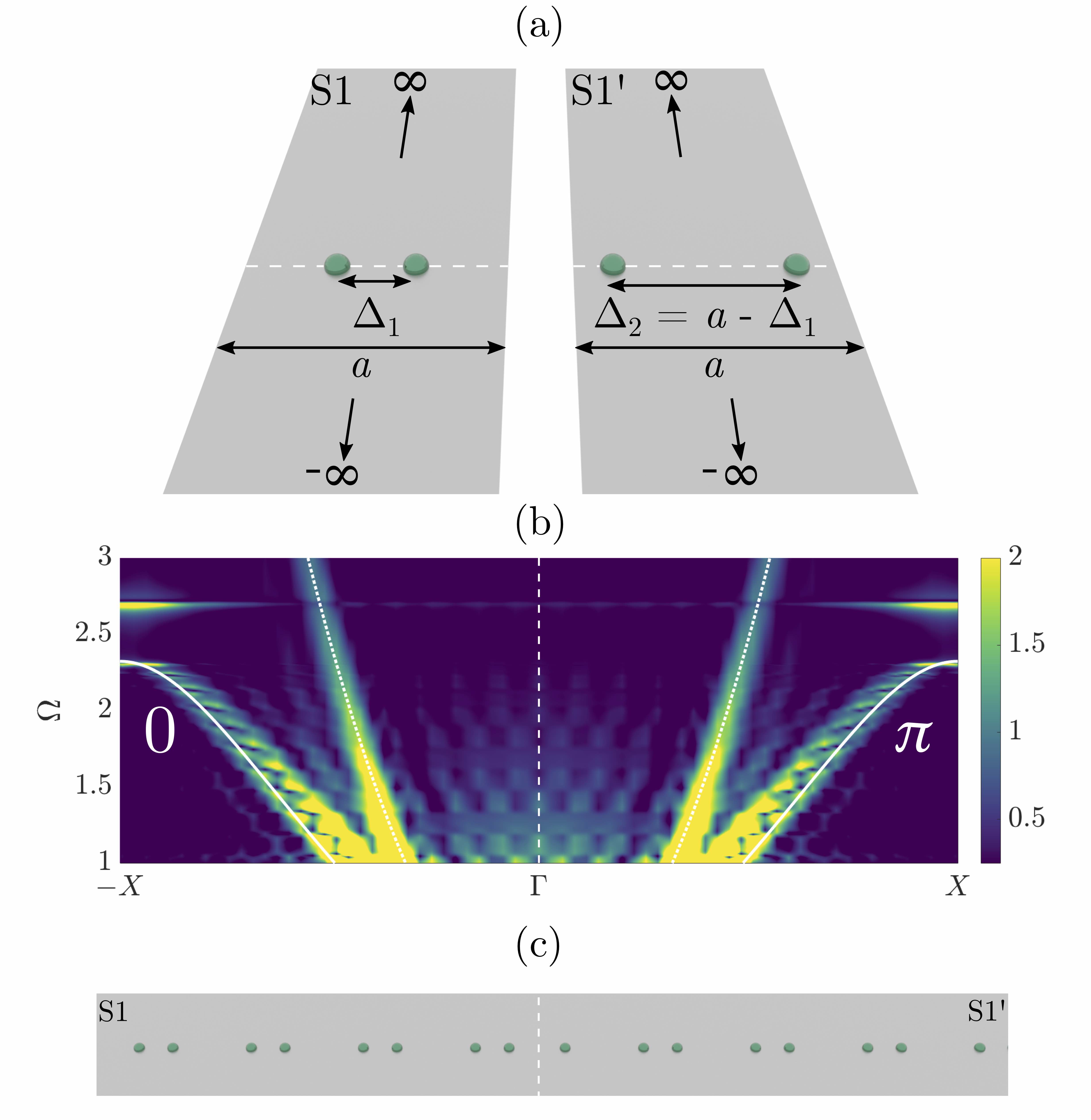}
     \caption{(a) Schematics of the infinite strips, that periodically repeat, that characterise the array and which are used for the dispersion curves for structures S1 and S1$^\prime$, such that $a = 1$, $M = 5$, $\Delta_{1} = 0.2$, $\Delta_{2} = a-\Delta_{1}$. The dispersion curves, from the spectral method \cite{chaplain2019rayleigh}, for the normalised frequency, $\Omega(\kappa)$ are shown in white, with the dotted white line showing the free space flexural `sound line', which, for KL plates is not dispersionless. Plotted along the wavenumbers from $\kappa =-X \equiv -\pi/a$ to $\kappa = \Gamma \equiv 0$ are the curves for S1, with $\Gamma$ to $\kappa = X \equiv \pi/a $ showing those for S1$^\prime$. The two dispersion relations are clearly identical. Further to this, the Fourier spectrum is also shown in (b), through a FFT of scattering simulations of the SSH geometry shown in (c). Corroborated by the calculation of the distinct Zak phases, which label the bands in (b), there is an edge mode within the bulk band gap, highlighted at $\Omega = 2.7$. The topological nature of a bandgap is determined by summation over the Zak phase of all the bands below this gap \cite{zak1985symmetry,xiao2015geometric}, having no dependence on the bands above it. As such we only show the lowest dispersion branch of this system.}
     \label{fig:Dispersion}
\end{figure}

Considering an infinite periodic line array of point masses, capable of supporting propagating Rayleigh-Bloch modes which exponentially decay perpendicularly to the array, allows the governing equation to be formulated as a generalised eigenvalue problem; we do so by partitioning the array and plate into periodic infinite strips and by formulating the wavefield, $w(\boldsymbol{x})$ as a combination of a Fourier series and a decaying basis, as in \cite{chaplain2019rayleigh}. Employing Floquet-Bloch conditions, and invoking orthogonality, then characterises the dispersion relation for an arbitrary periodic strip of width $a$. Adopting the nomenclature conventional with topological systems, the eigensolutions (wavefields) of this system of equations are then written as 
\begin{equation}
\vert w \rangle = \sum\limits_{n,m}W_{nm}\exp[{i(G_{n}-\kappa)x}]\psi_{m}\left(y\right),
\end{equation}
where, for integer $n$, $G = 2n\pi/a$ is a reciprocal lattice vector, $\kappa$ is the Bloch wavenumber and $\psi_{m}(y)$ is an exponentially decaying orthonormal Hermite function. The advantages of this approach allows a spectral Galerkin method to accurately and rapidly characterise the dispersion relation, an example of which is shown in Fig.~\ref{fig:Dispersion}.

An advantage of having such explicit solutions for the eigenstates, satisfying (1) with periodic modulation, is that they can be used to obtain key information from the bulk bands in the form of topological invariants, specifically the Zak phase. To demonstrate the efficiency of this, and the existence of 1D topological edge states we utilise the SSH model. This has been utilised in many systems for transport \cite{chen2014nonlinear, meier2016observation} and to identify the existence of edge modes \cite{zhu2018zak}. Here we wish to exploit this model to emphasise the features of 1D topological defect states, and how they can be used for energy harvesting.

To build the SSH model, in the setting of a 1D array of point loaded masses on a KL elastic plate, we first consider an infinite, periodic 1D array consisting of infinite unit strips of width $a$, with two masses of mass $M$ placed symmetrically about the strips origin a distance $\Delta_{1}$ apart; this system has the dispersion relation highlighted in Fig.~\ref{fig:Dispersion}. As before, this cell configuration will be labelled as structure S1. Due to the translational invariance present in the infinite structure, the same periodic structure can be built by a translation of the unit strip by a distance $a/2$.  In this new unit cell, the masses are separated symmetrically about the strip origin by $\Delta_{2} = a - \Delta_{1}$; this configuration has an identical dispersion relation to S1, and we denote the unit cell of this structure S1$^\prime$. Structures S1 and S1$^\prime$ can be seen in Fig.~\ref{fig:Dispersion}(a), with their calculated dispersion relation shown in Fig.~\ref{fig:Dispersion}(b) (as also confirmed by the Fourier spectrum obtained through scattering simulations). 

We then form an SSH array by creating a 1D chain composed of repeated cells of S1 and S1$^\prime$ which meet at an interface (Fig.~\ref{fig:Dispersion}(c)). To determine whether a topological edge mode exists at this interface, we calculate the Zak phase \cite{Zak1989} for each band defined by S1 and S1$^\prime$; each material composing the SSH array has a common band-gap and, provided each periodic structure has a distinct Zak phase, the existence of an edge mode is guaranteed \cite{atala2013direct,Xiao2014,xiao2015geometric}. 

The Zak phase, $\varphi_{n}^{Zak}$, for the $n^{th}$ band is defined in terms of the Berry connection  $\mathcal{A(\boldsymbol{\kappa})}$ such that
 \begin{align}
     \varphi_{n}^{Zak} &= \int\limits_{BZ}\mathcal{A}(\bfkappa)d\bfkappa, 
     \label{eq:zak}
 \end{align}
 with 
 \begin{align}
     \mathcal{A}(\bfkappa) &= i\langle u_{\bfkappa}|\partial_{\bfkappa}u_{\bfkappa}\rangle,
 \end{align}
where BZ denotes the Brillouin Zone; there are several efficient methods capable of calculating such invariants \cite{Soluyanov2011,Gresch2017}. We opt to dove-tail the eigensolutions obtained from the spectral method (2) to calculate the Zak phase for each band, by ensuring that $|u_{n,\boldsymbol{\kappa}}\rangle$ is cell periodic such that  $\vert w \rangle = e^{-i\boldsymbol{\kappa}\cdotp\boldsymbol{r}}|u_{n,\boldsymbol{\kappa}}\rangle$. In doing so the required quantities are readily available from the obtained eigensolutions. We evaluate \eqref{eq:zak} over the discretised BZ in $\kappa$-space such that
 \begin{align}
     \begin{split}
         \int\limits_{BZ}\mathcal{A}(\bfkappa)d\bfkappa \rightarrow \sum\limits_{\bfkappa_{j}}d\bfkappa\langle u_{\bfkappa}|\partial_{\bfkappa}u_{\bfkappa}\rangle\big\vert_{\bfkappa=\bfkappa_{j}}, 
     \end{split}
 \end{align}
 resulting in
 \begin{align}
 \begin{split}
     \varphi_{n}^{Zak} &= -\mathfrak{Im}\left(\log\prod\limits_{j=1}^{J}\langle u_{n,\bfkappa_{j}}|u_{n,\bfkappa_{j+1}}\rangle\right).
 \end{split}
 \end{align}
The periodic gauge condition is satisfied through $|u_{n,\bfkappa_{J+1}}\rangle = e^{-i\mathbf{G}\cdotp\mathbf{r}}|u_{n,\bfkappa_{1}}\rangle$. Due to the intrinsic connection with Wannier charge centers \cite{kohn1959analytic,kivelson1982wannier}, provided we have inversion symmetry with respect to the array axis, we are guaranteed a quantised Zak phase of $0$ or $\pi$; indeed this can be inferred from the symmetry properties of the band edge states \cite{xiao2015geometric}. In this setting, these correspond to the flexural displacement fields being localised to the centre or edges of the strip respectively. 

As expected, we find distinct Zak phases as highlighted for the lowest band in S1 and S1$^\prime$ in Fig.~\ref{fig:Dispersion}(b). As such at the interface between S1 and S1$^\prime$ we have an analogue to an incomplete Wannier state: there exists an edge mode. This confirmed through the Fourier spectrum shown in Fig.~\ref{fig:Dispersion}(b). 

To visualise the edge mode we make use of the attractive Green's function approach \cite{evans2007penetration,torrent13a} that can be employed to calculate the total wavefield, subject to forcings $F^{(j)}$ from $J$ masses. This can be evaluated quickly, obtaining
\begin{equation}
w(\boldsymbol{x}) = w_i(\boldsymbol{x}) + \sum_{j=1}^{J} F^{(j)} g\left(\Omega, |\boldsymbol{x}-\boldsymbol{x}^{(j)}|\right),
\end{equation} 
where $w_i({\bf x})$ is the incident field. Using the well-known Green's function \cite{evans2007penetration},
\(
g\left(\Omega, \rho\right) = ({i}/{8\Omega^2})
\left[H_0(\Omega \rho) - H_0(i\Omega \rho)\right]
\),
the unknown reaction terms $F^{(j)}$ come from the linear system
\begin{equation}
F^{(k)} = M^{(k)}\Omega^2\left[w_i({\boldsymbol{x}}^{(k)}) + \sum_{j=1}^{J}
  F^{(j)} g\left(\Omega, |{\boldsymbol{x}}^{(k)}-{\boldsymbol{x}}^{(j)}|\right) \right]. 
\end{equation} 
From this, fast Fourier transform (FFT) techniques can be utilised to obtain the dispersion relation in $\kappa$-space, shown in Fig.~\ref{fig:Dispersion}(b). Using this method, we demonstrate the characteristics of a 1D edge mode, by also evaluating the time-averaged flux through \cite{norris1995scattering}
\begin{equation}
    \langle \boldsymbol{F} \rangle = \frac{\Omega}{2}\mathfrak{Im}\left(w(\boldsymbol{x})\nabla^{3}w^{*}(\boldsymbol{x}) -\nabla^2w^{*}(\boldsymbol{x})\nabla w(\boldsymbol{x})\right).
\end{equation}

\begin{figure}[h!]
    \centering
    \includegraphics[width = 0.95\textwidth]{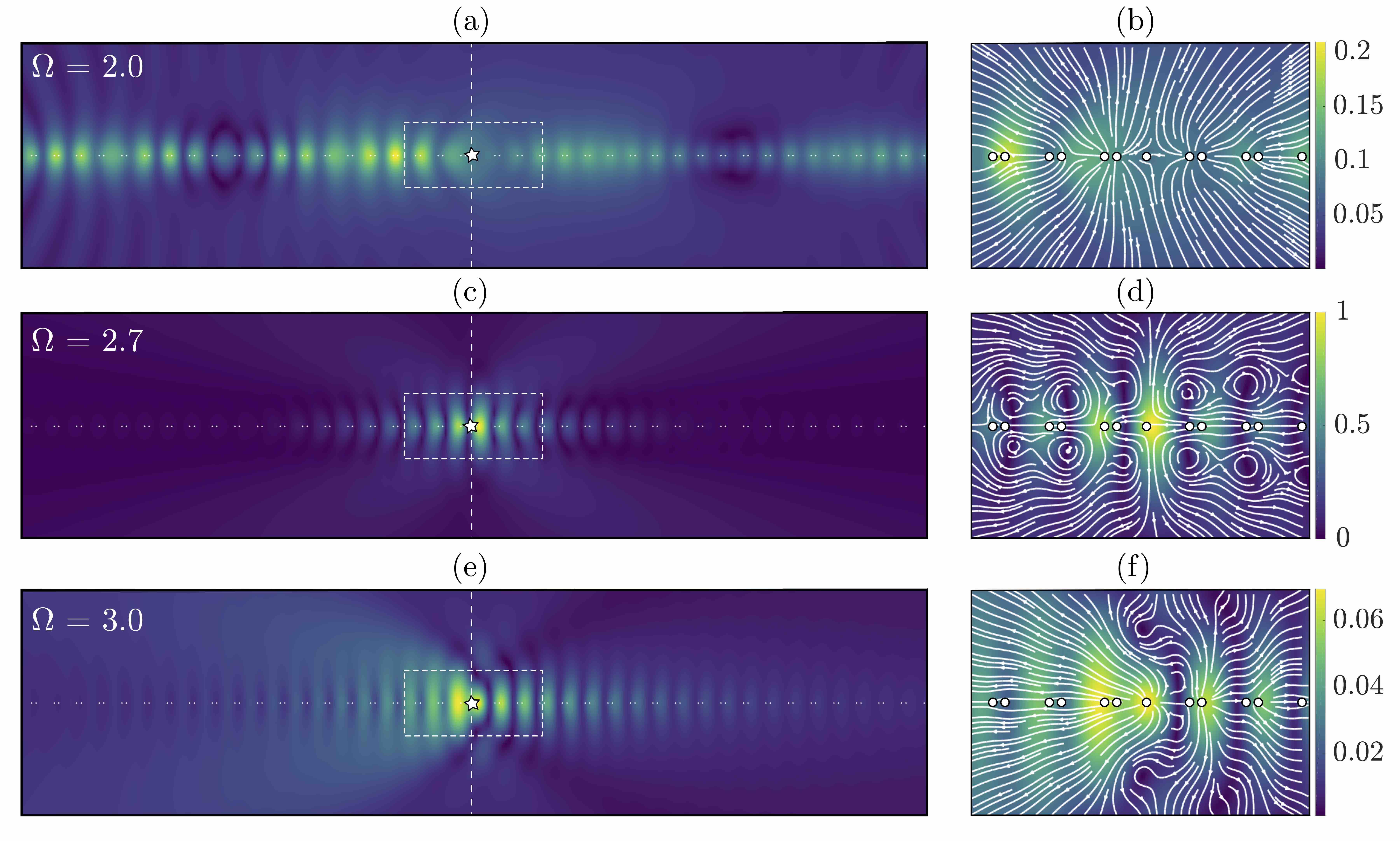}
    \caption{Scattered fields for (a) propagating, (c) edge mode and (e) conventional scattering defect modes, excited at the interface of S1 and S1$^\prime$, marked by the white stars. The field amplitudes are normalised with respect to the maximum amplitude of the scattered field of the topological edge mode, showing that it is approximately 17 times greater than the maximum amplitude of conventional defect modes. Panels (b,d,f) show a streamline plot of the time averaged flux in the regions highlighted by the rectangular box in (a,c,e) for the propagating, edge and defect modes respectively. The chiral nature of the edge mode flux is markedly different from the other cases.}
    \label{fig:TopoFlux}
\end{figure}
\begin{figure}[h!]
    \centering
    \includegraphics[width = 0.95\textwidth]{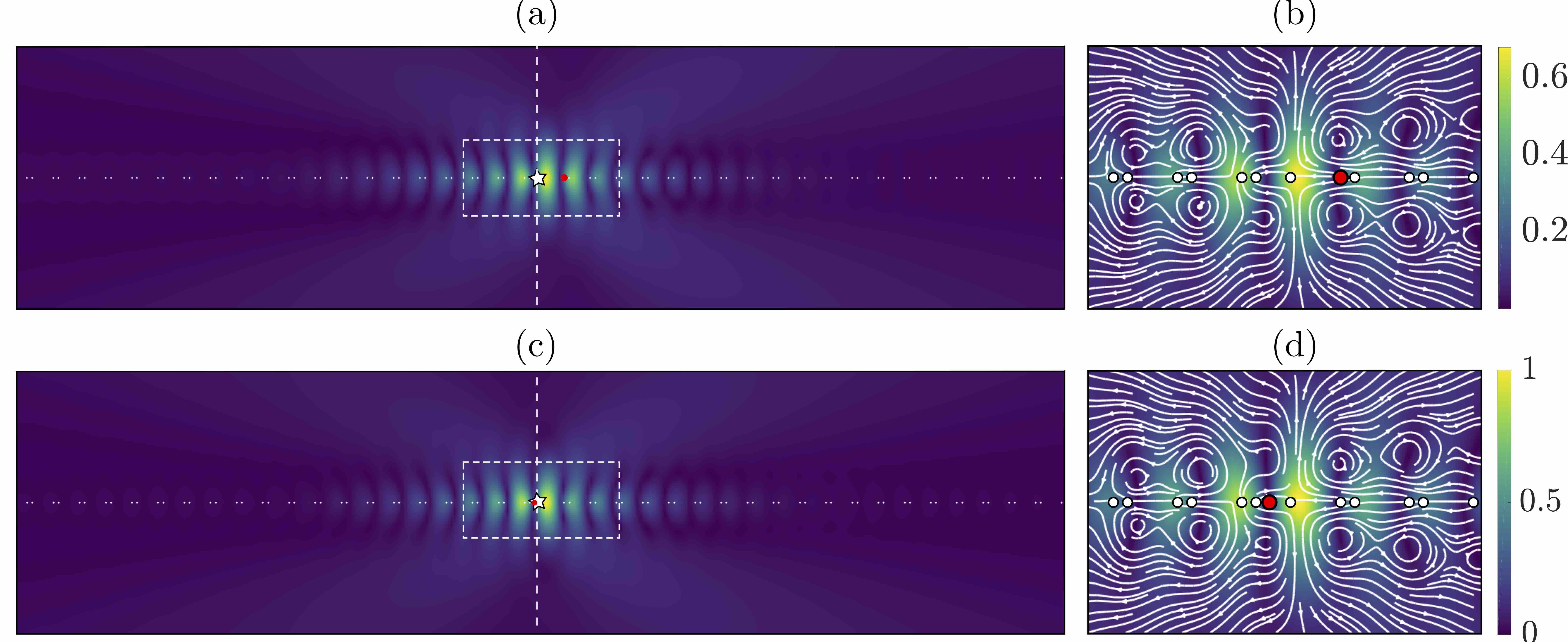}
    \caption{Testing disorder in the system, with (a-b) showing scattered fields and flux for a line-defect structure; a mass is removed in the first cell of S1$^\prime$ to the right of the interface (highlighted by the red point). (c-d) shows similar plots for an impurity-type defect; the red point represents an additional mass of $M = 1$ to the left of the source. In each case, the field is normalised with respect to the maximum amplitude of the perfect SSH case (Fig.~\ref{fig:TopoFlux}(c)). The chirality of the flux is seen to be preserved.}
    \label{fig:DisorderFlux}
\end{figure}

Shown in Fig.~\ref{fig:TopoFlux} are the scattered fields, for a monopolar point source placed at the interface between structures S1 and S1$^\prime$ in the SSH model, for the parameters as defined in Fig.~\ref{fig:Dispersion}. Exciting at different frequencies reveals three distinct modes present in the system: propagating, conventional defect and topologically protected edge modes. When exciting at $\Omega = 2$, unsurprisingly a propagating Rayleigh-Bloch mode exists, transiting along the array in each direction. Increasing the frequency to lie within the band gap, we see stark contrasts between the wavefields between the edge mode ($\Omega = 2.7$, Fig.~\ref{fig:TopoFlux}(c-d)) and a conventional localised defect state ($\Omega = 3$ Fig.~\ref{fig:TopoFlux}(e-f)); the amplitude of the topologically protected edge state is nearly 17 times that of the localised defect state, with its flux displaying chiral orbits which are indicative of edge modes, induced by the distinct topological phases at the interface \cite{makwana2018designing}. We further test the robustness of the topological edge state, by introduced line and impurity defects, by the removal and addition of extra masses respectively, demonstrated in Fig.~\ref{fig:DisorderFlux}: in each case the amplitude and chirality of the fields are preserved. 

We have successfully shown that the SSH model can be implemented in the setting of point mass loaded KL elastic plates. The existence of edge modes is confirmed through a variety of numerical techniques. The purpose of exploring such features in this system is to motivate energy harvesting applications in elastic settings, particularly because the localised amplitudes of edge modes are so much greater than those for conventional defect modes. A key feature of such harvesting structures is the ability to recycle energy from a distance; until now we have only focused on source positions at the interface between topologically distinct media. In order to assess the feasibility of harvesting devices, we explore the excitation of this mode from a distance. 

\begin{figure}[h!]
 \centering
    \includegraphics[width = 0.95\textwidth]{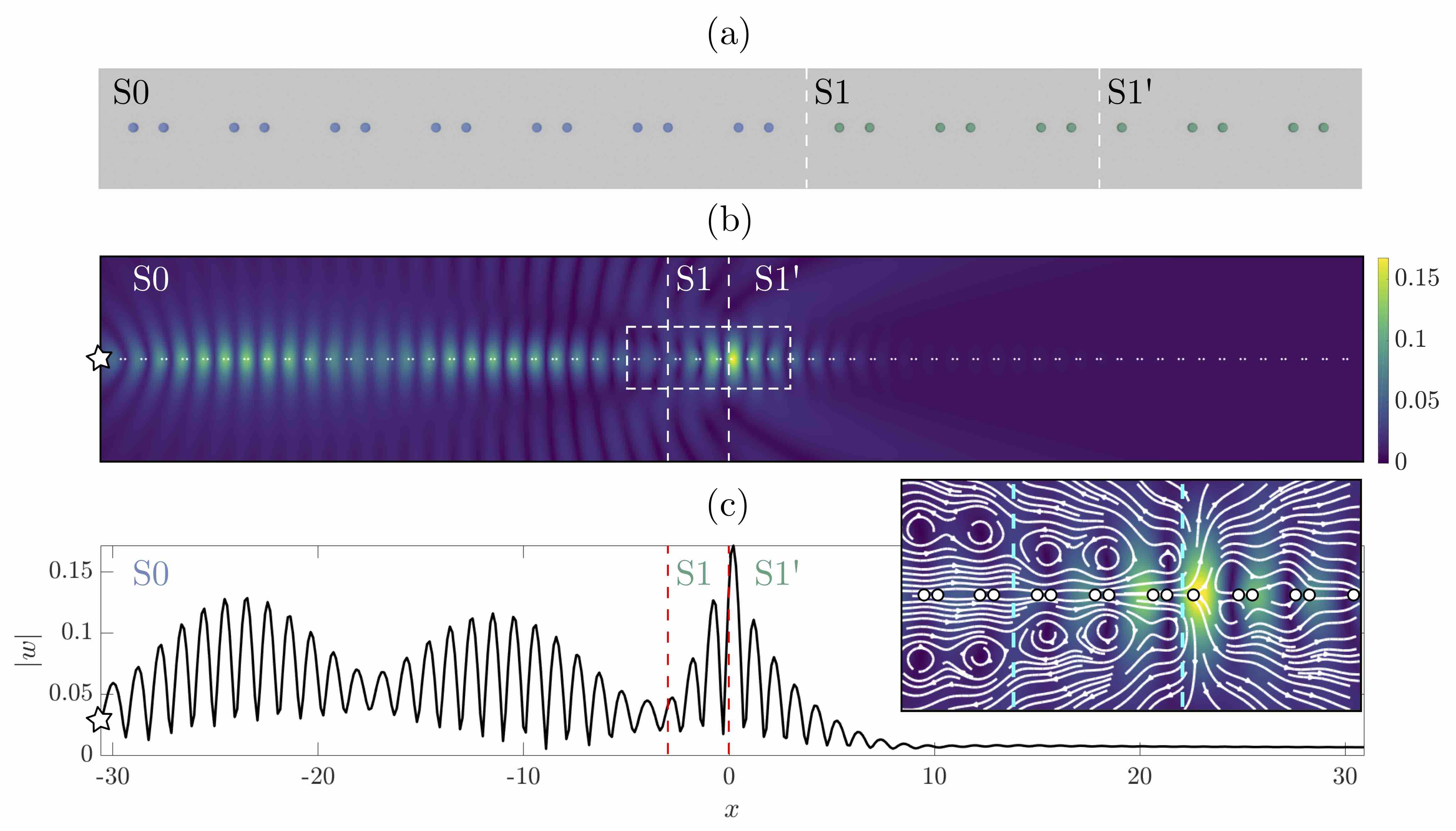}
    \caption{Exciting an edge mode using a step `grading'. (a) shows a schematic of the array, composed of structures S0 of $M = 2.3$ (blue masses), three cells of structure S1 (green masses) and finally a region of S1$^\prime$, forming a step-SSH array. In region S0, the frequency $\Omega = 2.7$ corresponds to a slow propagating mode that, upon reaching S1, excites a decaying mode in S1. This mode then feeds into the interface region between S1 and S1$^\prime$; at this frequency an edge mode is excited, shown by the chiral fields in the inset. (b) shows the scattered field, normalised to the amplitude of the perfect SSH array (Fig.~\ref{fig:TopoFlux}(c)). (c) shows the absolute amplitude $\vert w \vert$ along the array axis, indicating the increased amplitude at the interface between S1 and S1$^\prime$, demonstrating this edge mode can be externally excited.}
    \label{fig:ExternalExcitation}
\end{figure}

To do this, we consider a region, S0, consisting of the same geometric structure as S1, but with a lower mass value ($M = 2.3$) such that a propagating mode exists at the frequency of the edge mode in the SSH model. Then, at a given spatial position, we abruptly switch the mass value to be consistent with S1 ($M = 5$) - in this region an exponentially decaying mode is excited. The SSH interface is then encountered, by constructing a region of S1$^\prime$ close to the interface between S0 and S1. A schematic of this is shown, along with the field and flux computations in Fig.~\ref{fig:ExternalExcitation}, showing that it is possible to excite this mode from a distance; the amplitude of the resulting mode depends on the decay length introduced in the transition region between S1 and S1$^\prime$, a feature which can be predicted from high frequency homogenisation techniques \cite{chaplain2019rayleigh}. Thus 1D edge modes can be excited by a source which is external to the topological interface and this motivates energy harvesting applications within such regions. Limitations of this simplistic system are, however, immediately apparent; there must be a propagating region before the interface, and the effect is extremely narrowband. To circumnavigate these deficiencies, we turn to recent metawedge structures, and hybridise the SSH model with an adiabatic grading in a system of resonant rods atop an elastic beam.

\section{The Graded SSH metawedge}

The graded resonant metawedge \cite{colombi2016seismic} has proved an important source of inspiration for the `trapping' of energy, by a reduction in effective group velocity of propagating waves. The classical arrangement is that of resonant rods atop an elastic half-space, or elastic beam, as shown in Fig.~\ref{fig:IntroFig}(b). In this example, the rods adiabatically change in height from one unit cell to the next, generating locally periodic cells; the global behaviour of the device is inferred from the dispersion curves corresponding to an infinite array of each rod height \cite{Romero-Garcia2013}. As such, different frequency components encounter local bandgaps at different spatial positions. Similar to rainbow trapping devices \cite{Tsakmakidis2007}, the metawedge achieves local field enhancement which can be used for energy harvesting effects \cite{DePonti2019}. Despite the success, both in design and experimental verification, of a wide variety of effects exhibited by the metawedge and similar structures \cite{Colombi2017,skelton2018multi,chaplain2019tailored}, this simplistic array has reflections, due to Bragg scattering, at the `trapping' positions. As such, energy is not confined for prolonged periods due to intermodal coupling, and rainbow reflection phenomena is seen instead \cite{chaplain2020rainbow}.


Topological systems therefore seem attractive candidates for energy extraction, due to their resilience to back-scatter and strong confinement; the longer energy is confined to a spatial position, the more energy that can be harvested \cite{chaplain2020rainbow}. This is more efficient for symmetry broken systems, where a lack of coupling to reflected waves leads the energy to be more localised; a natural extension of this is to consider topological devices. Indeed, recent designs for topological rainbow effects have been theorised for elasticity in perforated elastic plates of varying thickness, based  on topologically protected zero-line-modes (ZLMs) between an interface of 2D square array structures \cite{ungureanu2020route,Makwana2019tuneable}. 

Due to the low dimensionality of the 1D-SSH system, the SSH model provides an optimal arrangement for elastic energy harvesting as there is no propagating component of the edge mode. However, the caveat to this has already been alluded to - this mode only exists for a very narrow range of frequencies. We therefore design a hybrid graded-SSH-metawedge which, based on a gentle adiabatic grading of alternating SSH structures, significantly increases the bandwidth of operation, serving as the perfect candidate for topological rainbow trapping. Due to the strong interaction between the symmetry broken structure and the edge mode, these devices offer an additional benefit of being compact compared to classical metawedges.

\begin{figure}[H]
    \centering
    \includegraphics[width = 0.8\textwidth]{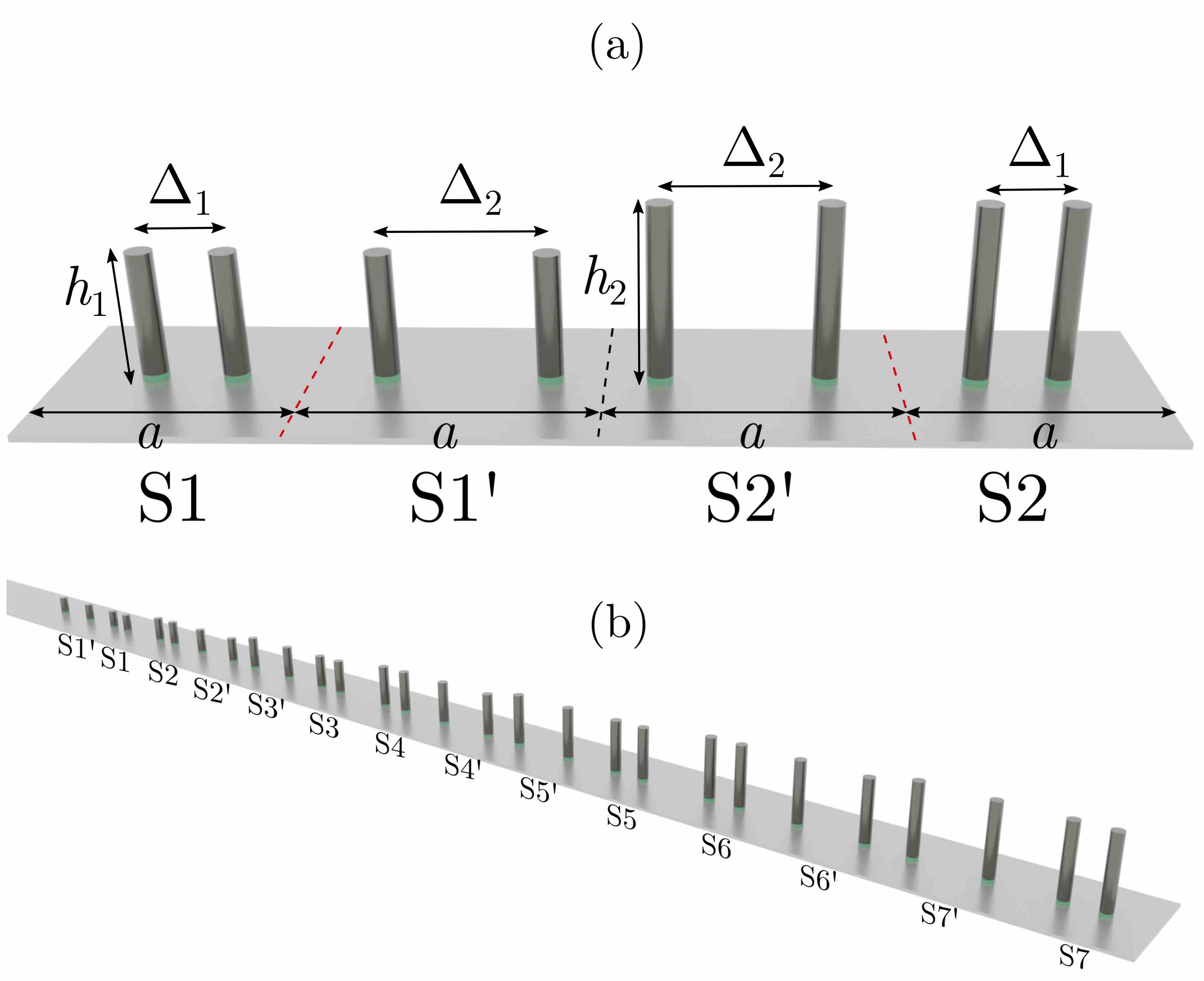}
    \caption{Graded SSH schematics: (a) shows the Sn-Sn$^\prime$-Sm$^\prime$-Sm altering cell structure for heights $h_{n}, h_{m}$ (b) shows graded-SSH-metawedge for 7 alternating SSH cells. The green disks at the base of each rod represent the positioning of the piezoelectric material discussed in Section~\ref{sec:Harvesting}.}
    \label{fig:GradedSSH}
\end{figure}

Figure~\ref{fig:GradedSSH} shows the proposed design, as a zoom of Fig.~\ref{fig:IntroFig}, consisting of resonant rods atop an elastic beam. Similar to the motivational mass loaded case, we define structures S1 and S1$^\prime$ to be unit cells consisting of rods of height $h_{1}$ arranged in the SSH configuration. The heights of the rods are adiabatically increased every two unit cells, with the arrangement being mirrored: cells with rods of height $h_{2}$ follow an S2$^\prime$-S2 interface: this is repeated along the array. An example of the corresponding Sn-Sn$^\prime$-Sm$^\prime$-Sm geometry (where $n$ and $m$ corresponding to heights $h_{n}, h_{m}$) is shown in Fig.\ref{fig:GradedSSH}(b). Throughout the following sections the existence of the edge modes is confirmed for the now familiar S1-S1' configuration, followed by an investigation into the uses for elastic energy harvesting.

\section{Topological rainbow trapping for elastic energy harvesting}
\label{sec:Harvesting}
Topological systems have been widely proposed as efficient solutions for elastic energy transport, guiding and localization \cite{makwana2018designing,Makwana2019tuneable}.
These concepts offer, amongst others, promising capabilities for energy harvesting, due to the enhancement of local vibrational energy present in the environment. One of the main challenges in elastic energy scavenging, is obtaining simultaneously broadband and compact devices \cite{ErturkBook}. Broadband behaviour is usually achieved through nonlinear effects \cite{Nonlinear,ERTURK20112339} or multimodal response \cite{shahruz2006design,FERRARI2008329}, i.e. by exploiting multiple bending modes of continuous beams or arrays of cantilevers. Whilst multimodal harvesting enhances the operational bandwidth, it is usually accompanied by an increase in the volume or weight of the device. This can affect the overall power density of the system as well as the circuit interface, which becomes more complex with respect to single mode harvesters. Conversely it is important to appreciate that multimodal schemes can be well integrated with metamaterial concepts, leading to truly multifunctional designs \cite{Sugino} with enhanced energy harvesting capabilities. 
\begin{figure}[h!]
    \centering
    \includegraphics[width = 0.9\textwidth]{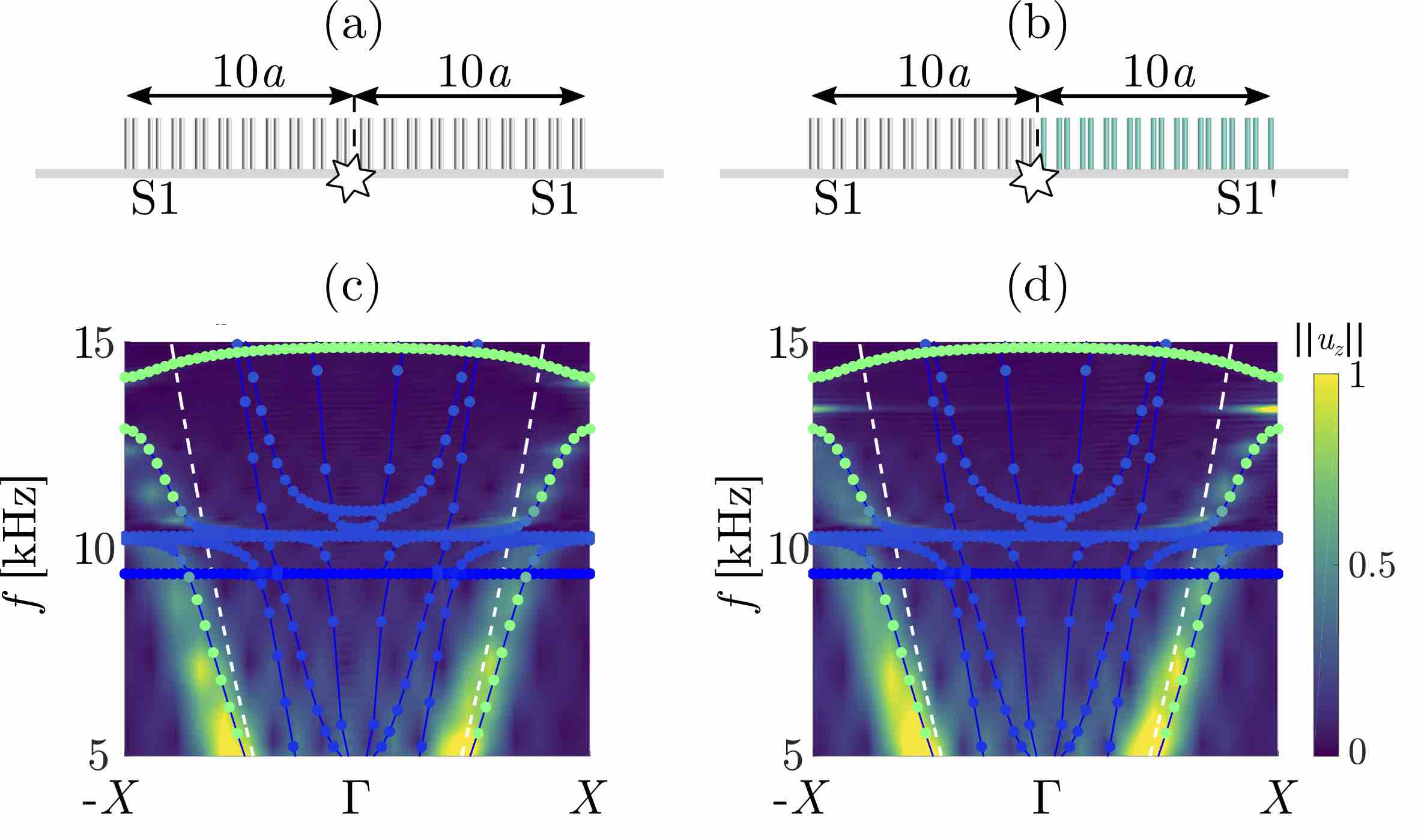}
    \caption{Dispersion comparisons for (a) trivial and (b) SSH interfaces. (a-b) show the arrangement of rods atop a beam with 10 cells of width $a$ consisting in (a) of structures S1-S1 (grey rods), forming a trivial interface, with (b) showing an SSH interface between 10 cells of structure S1 and 10 cells of S1' (green rods). (c-d) shows the Fourier spectrum for this arrangement from a scattering simulation where a source is placed at the interface, marked by the star and dashed black lines in (a-b). An edge mode appears inside the bandgap defined by the longitudinal resonance of the rods, as expected in the SSH arrangement. Overlaid in (c-d) are the numerical dispersion curves for a perfectly periodic, infinite array of structures S1 (and simultaneously S1$^\prime$) represented by the coloured points in (c-d), with green points corresponding to vertical polarization of the rod (axial elongation), whilst blue refers to horizontal (flexural motion). The geometrical parameters are such that a = 30 $ \si{\milli\meter}$ with the SSH spacing $\Delta_{1} = 10$ $\si{\milli\meter}$. The rods have a height of $h = 82$ $\si{\milli\meter}$ and circular cross section of radius $r = 3$ $\si{\milli\meter}$. The beam has thickness $t=10$ $\si{\milli\meter}$ and width $w = 30$ $\si{\milli\meter}$, as is assumed to be infinitely long in the direction of the array. }
    \label{fig:DispersionRod}
\end{figure}

Here we propose a multimodal scheme, i.e. a broadband device, which is simultaneously compact due to the reduced number of required cells. The device is similar to that in \cite{DePonti2019}, but based on the excitation of local edge modes through the graded-SSH-metawedge geometry (Fig.\ref{fig:GradedSSH}). We recall that the physics of these arrays is primarily governed by the longitudinal (axial) resonances of the rods \cite{colombi2016seismic} which, along with the periodicity, determine band-gap positions through their resonance. The axial resonance frequency of the rod is governed by the rod height \cite{colombi2016seismic}. By a simple  variation of the length of adjacent rods, an effective band-gap, that is both broad and sub-wavelength can be achieved. The addition of alternating SSH configurations introduces frequency dependent positions of localized edge states. By the definition of rainbow effects \cite{chaplain2020rainbow}, this will hence define a true topological rainbow.

To quantify the advantages of such designs for energy harvesting, we compare its performance with a conventional rainbow reflection device \cite{DePonti2019} composed of equal number of rods, with identical grading angle and quantity of piezoelectric material. 

We firstly verify the existence of an edge mode, by considering two arrays, one only composed of equal rods with constant spacing, i.e. consisting only of structures S1 (Fig.~\ref{fig:DispersionRod}(a)), and another with a transition between regions consisting of structures S1 and S1$^\prime$, shown in Fig.~\ref{fig:DispersionRod}(b), similar to the previous examples (Figs.~\ref{fig:IntroFig}(a),~\ref{fig:TopoFlux}). Both systems are made of aluminium ($\rho=2710~\si{\kilo\gram\meter^{-3}}$, $E=70~\si{\giga\pascal}$ and $\nu=0.33$) and composed of rods with length $82~\si{\milli\meter}$ and circular cross section with $3~\si{\milli\meter}$ radius. The beam is defined by $10~\si{\milli\meter}$ thickness and $30~\si{\milli\meter}$ width, and is assumed to be infinitely long in the direction of the wave propagation. The unit cell dimension is $a=30~\si{\milli\meter}$, with the resonator separation inside the cell as $\Delta_1=10~\si{\milli\meter}$ (in structure S1) and $\Delta_2=a-\Delta_1 =20~\si{\milli\meter}$ (in S1$^\prime$). We compute the dispersion curves for both configurations using Abaqus \cite{abaqus2011abaqus} with a user defined code able to impose Bloch-Floquet boundary conditions. As expected, the two dispersion relations are identical as can be seen from the directions $-X-\Gamma$ and $\Gamma-X$ in Fig.~\ref{fig:DispersionRod}(b). To detect the presence of an edge mode, we excite both systems with a time domain frequency sweep in the range $5-15~\si{\kilo\hertz}$, with a source inside the array and located at the interface between S1-S1$^\prime$. By inspection of the spatiotemporal Fourier transform of the resultant wavefield, an edge mode clearly appears inside the bandgap opened by an axial resonance. This can be seen through the colormap of the dispersion curves where the polarisation is such that green points correspond to vertical (longitudinal) polarisation, whilst blue refers to horizontal (flexural) polarisation (see Fig.~\ref{fig:DispersionRod}(c,d)). 

 \begin{figure}[h!]
    \centering
    \includegraphics[width = 0.95\textwidth]{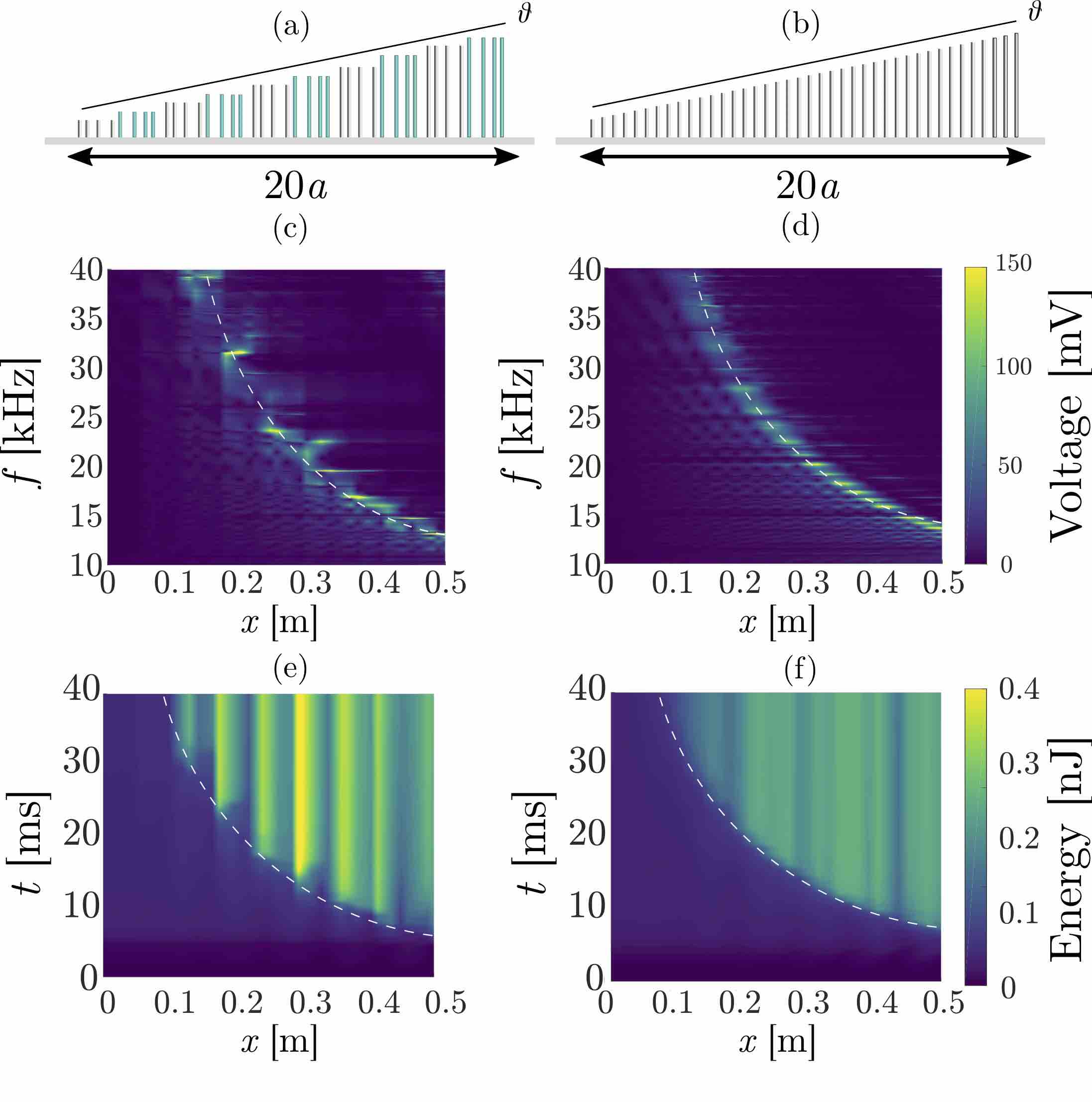}
    \caption{(a)-(b) Schematics of graded-SSH-metawedge and conventional metawedge respectively. Open circuit voltage (c)-(d) and accumulated energy (e)-(f) for the graded SSH and conventional metawedges as a function of position along the array.}
    \label{fig:VoltageEnergy}
\end{figure}
We consider graded line arrays of resonators, to simultaneously excite the array from outside and to enlarge the bandwidth, based around the designs shown in Figures~\ref{fig:GradedSSH} and \ref{fig:VoltageEnergy}(a).
 To quantify the energy that can be stored, we insert PZT-5H piezoelectric disks ($\rho=7800~\si{\kilo\gram\meter^{-3}}$, $E=61~\si{\giga\pascal}$ and $\nu=0.31$) of $2~\si{\milli\meter}$ thickness between the rods and the beam (shown as green disks in Fig.~\ref{fig:GradedSSH}). Due to the dominant axial elongation in the rod response, we model the piezoelectric coupling by means of the $33$ mode piezoelectric coefficient $e_{33} = 19.4~ \si{\coulomb\meter^{-2}}$,  and constant-stress dielectric constant $\epsilon_{33}^T/\epsilon_0=3500$, with $\epsilon_0 = 8.854~\si{\pico\farad\meter^{-1}}$ the free space permittivity. The device is composed of $40$ rods with height approximately from $5~\si{\milli\meter}$ to $100~\si{\milli\meter}$ and grading angle $\theta\simeq 4.7^{\circ}$. We compare the SSH rainbow system with a conventional rainbow device, through a steady state dynamic direct analysis performed using Abaqus with open circuit electric conditions. The infinite length of the beam is modeled using ALID boundaries at the edges \cite{RAJAGOPAL201230}. We see rainbow effects in both cases (Fig.~\ref{fig:VoltageEnergy}(c,d)), i.e. spatial signal separation depending on frequency, but the voltage peaks are more localized and with higher amplitude in the SSH case. It is important to notice that this effect is more significant in the steady state regime; a relatively long excitation is required in order to properly activate the edge modes. Both systems are compared using a time domain simulation with a frequency sweep in the range $10-40~\si{\kilo\hertz}$ with a source duration of $40~\si{\milli\second}$. In order to quantify the amount of electric energy stored in both cases, we attach each piezo disk to an electric load of $10~\si{\kilo}\Omega$ by means of a user Fortran subroutine integrated with Abaqus implicit time domain integration scheme. The accumulated energy as a function of time is shown in Fig.~\ref{fig:VoltageEnergy}(e,f). The excitation of the edge modes at discrete frequencies can clearly be seen, with an approximate maximum value of stored energy of $0.44~\si{\nano\joule}$. For the conventional metawedge, we see that energy is more evenly distributed along space, with a maximum value of approximately $0.26~\si{\nano\joule}$. This implies that, once the edge modes have been efficiently excited in the SSH configuration, we obtain a local enhancement of approximately $40\%$ of the trapped electric energy when compared to conventional reflective rainbow metawedge configurations.
 
 \begin{figure}
    \centering
    \includegraphics[width = 0.9\textwidth]{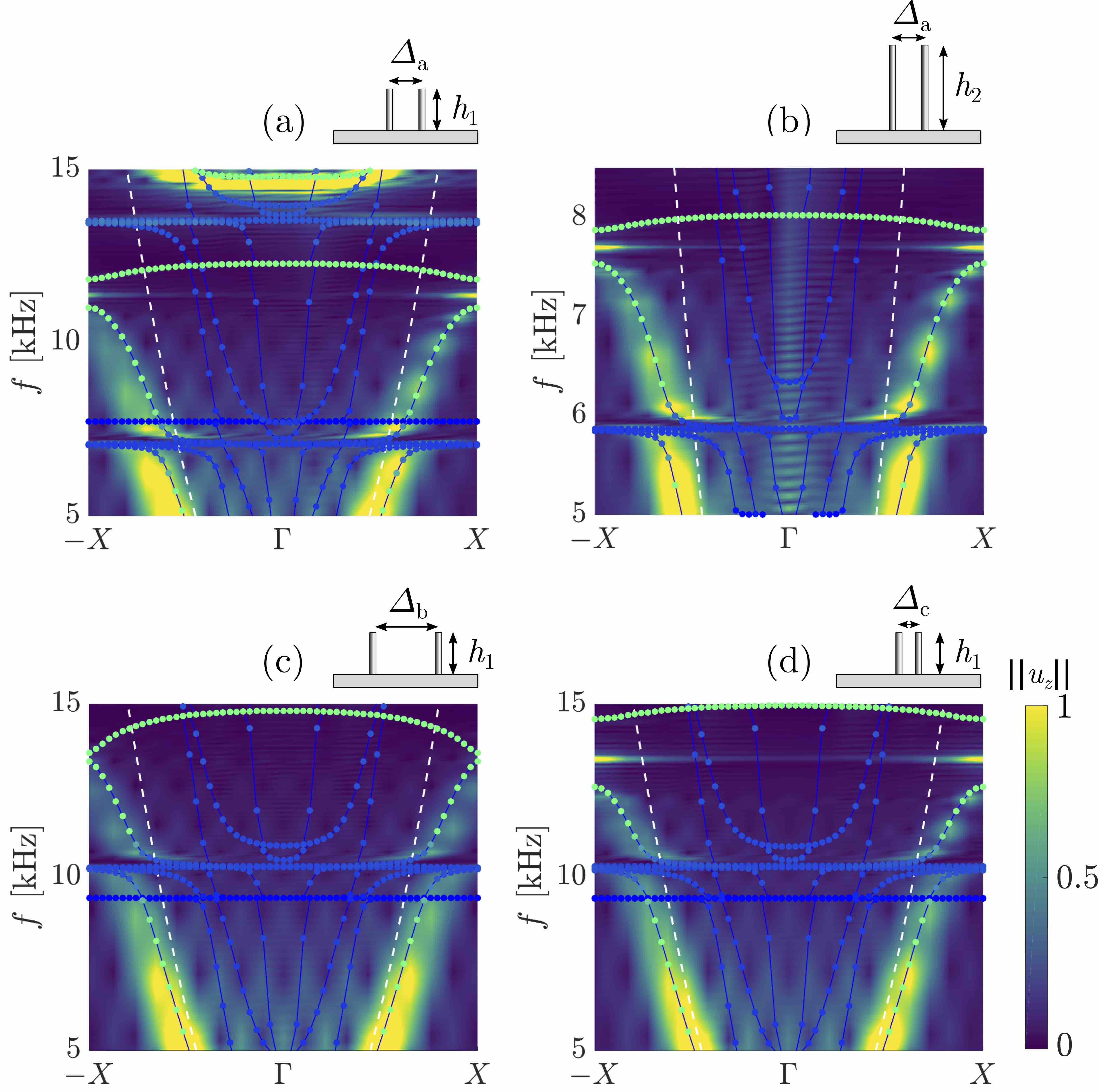}
    \caption{Grading height and spacing: Polarised dispersion curves, with longitudinal motion shown by green points, overlaid on the Fourier spectra of the SSH interfaces consisting of different parameters within S1 and S1$^\prime$. The periodicity $a = 30$ $\si{\milli\meter}$ and rod radius $r = 3$ $\si{\milli\meter}$ remain constant throughout. Panels (a,b) show the effect of grading the height of the resonators; the longitudinal resonances of the rods influence the position of the Bragg gap, allowing for increased bandwidth of the device. (a) has parameters such that $\Delta_{1} = \Delta_{a} =  10$ $\si{\milli\meter}$, $h_{1} = 100$ $\si{\milli\meter}$ whilst (b) has the same S1 spacing with $h_{2} = 155$ $\si{\milli\meter}$. The edge modes are clearly visible. Note the different scaling on the frequency axis; the gap position has been decreased due to the longitudinal resonance of the rods. Panels (c,d) show the effect of grading the spacing $\Delta_{1}$ in structures S1. There is a much smaller range of values this can take as we are limited by the symmetry of the unit cell. Panel (c) shows the band gap opening for a large $\Delta_{1} = \Delta_{b} = 14$ $\si{\milli\meter}$ and (d) shows the edge mode for $\Delta_{1} = \Delta_{c} = 7$ $\si{\milli\meter}$.}
    \label{fig:Height_Spacing}
\end{figure}

Before finalising this design in 3D elastic half-spaces, we address the question of why altering the height of the resonators was chosen as the grading parameter other than, say, the initial spacing $\Delta_{1}$ in S1. Figure~\ref{fig:Height_Spacing} shows the rationale behind this, which is to ultimately achieve broadband performance. Figures~\ref{fig:Height_Spacing}(a,b) show the effects of the longitudinal resonance on the position of the Bragg gap in which the edge mode lies; the taller the rod the lower in frequency the axial resonance, which pushes down in frequency the Bragg gaps below it; the geometrically induced bandgaps are influenced by the resonance frequencies of the rods, which permits a natural tunability of the devices. Exploiting this, as we do, by the adiabatic grading of the rod heights increases the range of frequencies at which the distinct edge modes exist and which can be therefore be exploited  for harvesting as shown in Figure~\ref{fig:VoltageEnergy}. Contrary to this, if instead we chose to alter the spacing $\Delta_{1}$, a much smaller range of edge mode frequencies can be exploited. Given the previous definitions of S1, we see that $\Delta_{1} < a/2$; for $\Delta_{1} > a/2$ structure S1 is simply interchanged with the primed configuration due to the symmetries of the unit cell, whereas if $\Delta_{1} = a/2$ it is not possible to create an interface which distinguishes between the two geometries. As such, there is a narrow range of values which $\Delta_{1}$ can take. This is highlighted in Figures~\ref{fig:Height_Spacing}(c,d), where the spacing $\Delta_{i}$ is marked with Roman subscripts to avoid confusion between $\Delta_{1}$ and $\Delta_{2}$ used in the definition of S1 and S1$^\prime$. In each case the different $\Delta_{i}$ correspond to altering $\Delta_{1}$. We alter from $\Delta_{b} = 14$ $\si{\milli\meter}$, which is close to the largest separation where the geometries can be distinguished, in Fig.~\ref{fig:Height_Spacing}(c) to $\Delta_{c} = 7$ $\si{\milli\meter}$ in Fig.~\ref{fig:Height_Spacing}(d). These show that the position of the bandgap, and hence edge mode frequency, is largely unaffected by the change in spacing. As such for the simplest designs, where there is only one grading parameter, the choice of grading the rod height leads to optimal performance.

\section{Graded-SSH-metawedge for Rayleigh waves}

Galvanised by the amplifications achieved by the 1D topological edge states in elastic beams, we now turn our focus to full 3D (isotropic) elastic half-spaces, patterned with arrays of resonant rods on the surface. This structuration creates a so-called metawedge and these have been used to exhibit extraordinary control of surface Rayleigh waves in terms of rainbow devices and tailored surface to body wave converters \cite{colombi2016seismic,Colombi2017,chaplain2019tailored}; here we explore graded SSH structures for the elastic half-space.

\begin{figure}[h!]
    \centering
    \includegraphics[width = 0.8\textwidth]{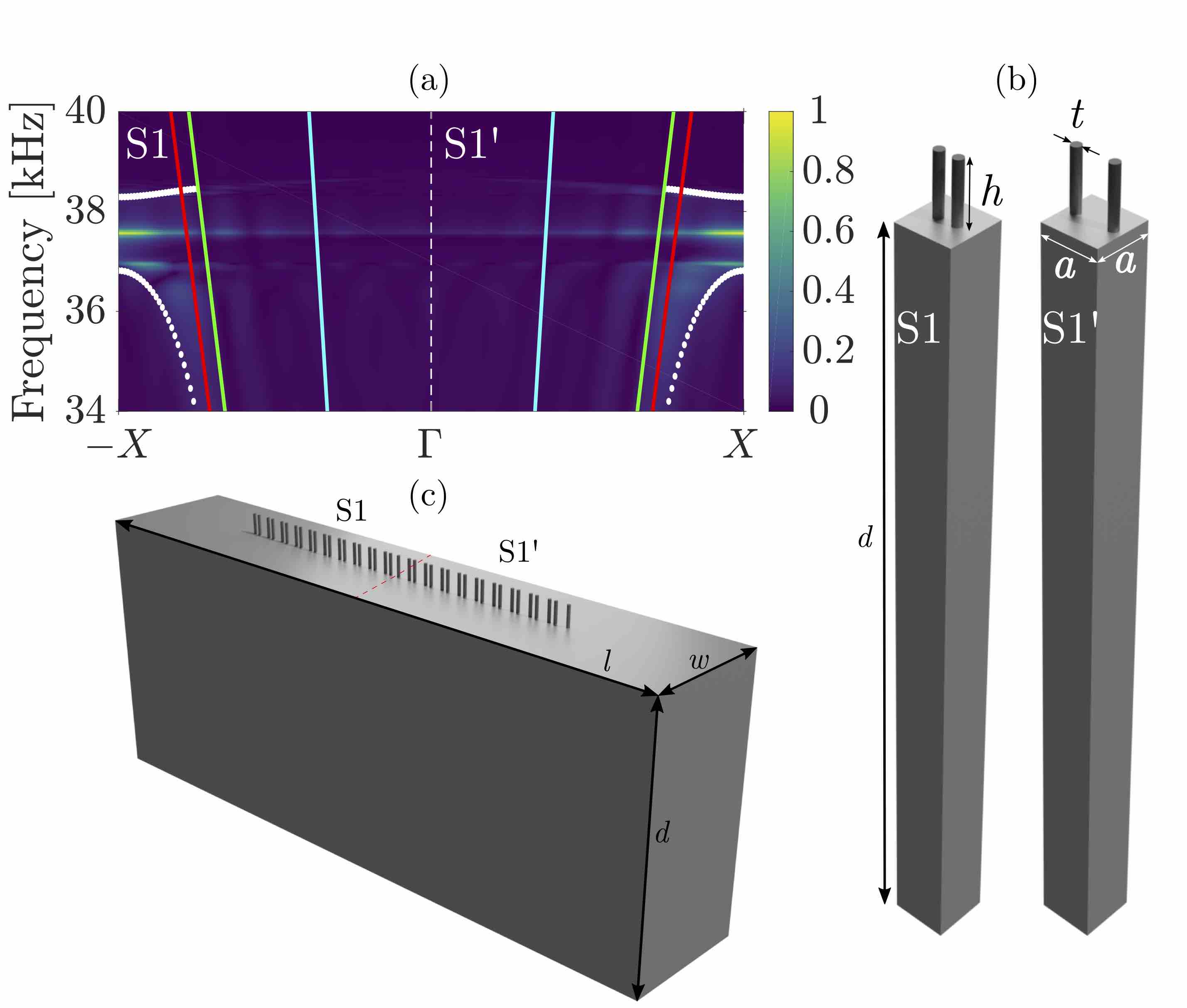}
    \caption{Panel (a) Shows identical dispersion curves 
    obtained using Comsol Multiphysics for S1 and S1$^\prime$, using the geometries defined in (b) as white points, with Rayleigh, shear and compressional sound lines shown in red, green and cyan respectively. Also shown, through the Fourier spectrum is the existence of an edge mode at the band gap centre. The source is placed at the interface between regions of structures S1 and S1$^\prime$ (each comprising 10 cells), marked by the dashed red line in (c). The parameters of the unit cells and half-space are such that are $a = 30~\si{\milli\meter}, h = 30~\si{\milli\meter}, t = 0.5~\si{\milli\meter}, l = 1~\si{\meter}, w = 20~\si{\centi\meter}, d = 40~\si{\centi\meter}$.}  
    \label{fig:HSDispersion}
\end{figure}

\begin{figure}[h!]
    \centering
    \includegraphics[width = \textwidth]{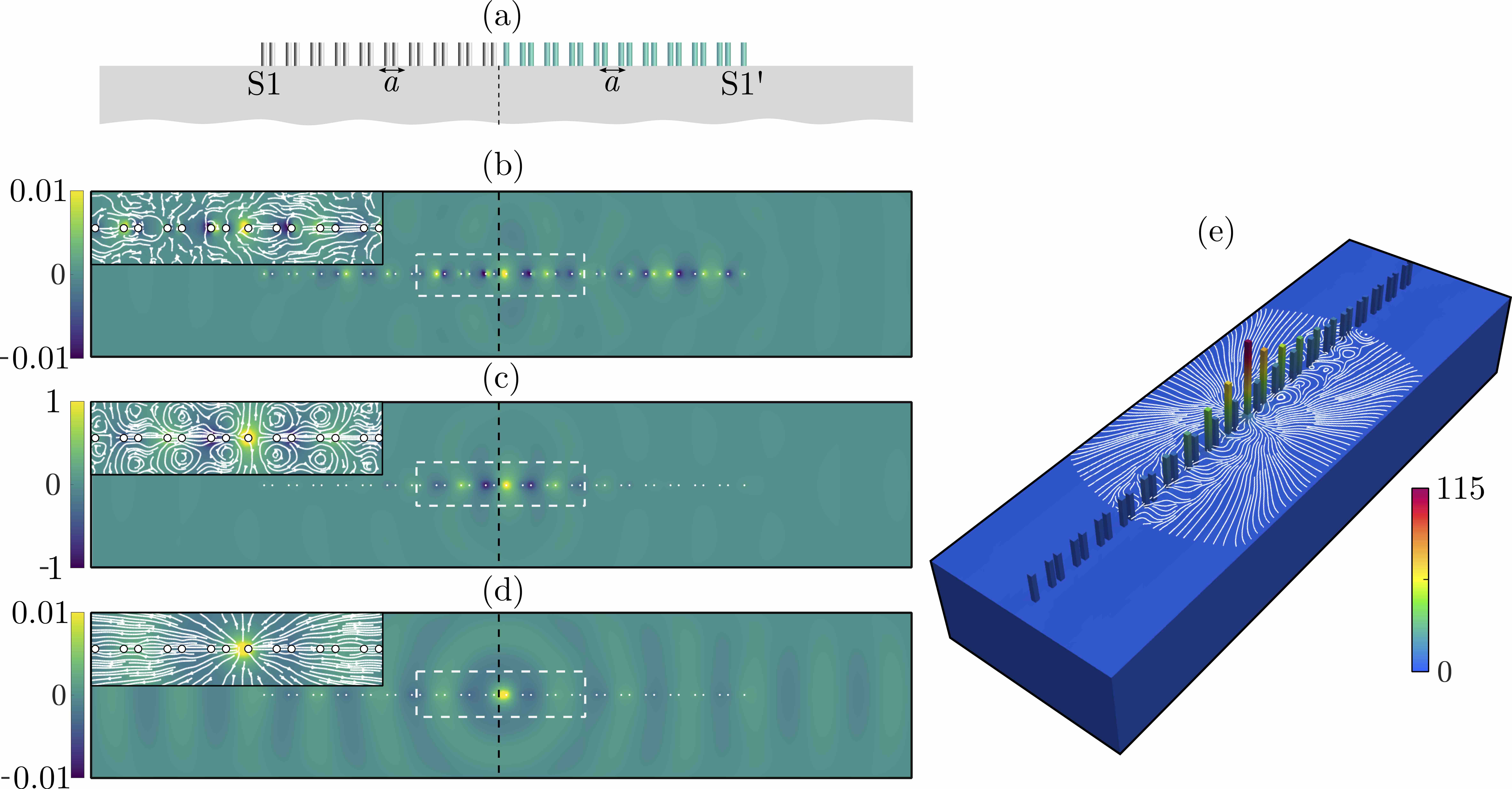}
    \caption{(a) Schematic of SSH interface, as a cross section of that in Fig.\ref{fig:HSDispersion}(c).  Regions composed of S1 are shown as grey rods with S1$^\prime$ shown as green. A top view of the surface displacement of the halfspace shows a propagating, edge and scattering mode in (b,c,d) respectively, normalised to the displacement of the edge mode. Their respective fluxes, within the dashed white rectangles, are shown in the insets. The arrays are forced at the SSH interface (dashed black line). Again the difference in amplitude (with the edge mode having approximately 100 times the amplitude) and flux patterns demonstrate the existence of the protected edge state. (e) Shows the vertical displacement of the rods relative to the maximum displacement of the surface in (c), for a frequency $\Omega = 37.7$ $\si{\kilo\hertz}$, near the edge mode frequency, excited at the SSH interface. Also shown is the chiral flux pattern. The relative vertical displacement of the rods near the SSH interface is over 100 times larger than the maximum vertical displacement of the half-space surface.}  
    \label{fig:HalfSpaceABFlux}
\end{figure}

Elastic half-spaces support a wider variety of waves than the motivational KL plates or elastic beams. Surface Rayleigh waves propagate along the free surface, exponentially decaying with depth into the bulk, travelling at wavespeed $c_{r}$, independently of any periodic structuring. Further to this, two polarisations of body waves, namely compressional (P) and vertical/horizontal shear (SV and SH) waves exist, both travelling at differing wavespeeds $c_{p}$ and $c_{s}$ respectively such that $c_{p} > c_{s} > c_{r}$ \cite{achenbach2012wave}. Unlike recently designed mode conversion devices \cite{chaplain2019tailored}, we focus here on exciting topologically protected surface waves by utilising the now familiar SSH model of rods, but now placed atop such a half-space. Analysis of the dispersion curves of two structures S1 and S1$^\prime$ show the existence of an edge mode at the domain boundary between the two geometries, shown in Fig.~\ref{fig:HSDispersion}, and confirmed through scattering simulations in Fig.~\ref{fig:HalfSpaceABFlux}.

We obtain these results from time domain simulations of an array consisting of 10 cells of S1 and 10 cells of S1$^\prime$ atop an aluminium halfspace, as shown in Fig.~\ref{fig:HSDispersion}(c). These simulations are carried out with SPECFEM3D, an open-source code fully parallelised with MPI \cite{peter2011forward}. This software solves the full 3D elastic wave equation leveraging the spectral element method for space discretization, with an explicit finite difference scheme for time integration. In order to reduce potential spurious reflections and emulate unbounded wave propagation, we apply Stacey absorbing boundary conditions at the edges of the half-space. The top surface and the resonator boundaries are traction-free to guarantee the propagation of surface waves.

The system is initially excited with a broadband sweep source ranging from $10$ to 50 $\si{\kilo\hertz}$, polarised in the vertical direction ($z$). By double Fourier transforming the output signal in both space and time we obtain the dispersion curves as the Fourier spectrum in Fig.~\ref{fig:HSDispersion}(a), which are then validated by means of an eigenvalue analysis in Comsol Multiphysics 5.4. We can readily compute the band structure of an infinite (doubly) periodic array for the unit cells designed in Fig.~\ref{fig:HSDispersion}(b) with Bloch-Floquet boundary conditions applied on the side boundaries, and a low-reflecting boundary condition on the bottom surface of the computational domain to avoid spurious reflections. Since we are only considering normally incident excitation we only display the dispersion curves along $\Gamma-X$ \cite{chaplain2019tailored}, allowing the one-dimensional behaviour to be inferred.

After a thorough analysis of the dispersion relations we select the input frequencies for the scattering simulations in SPECFEM3D. Here we excite the array at the SSH interface  with a sinusoidal source corresponding to propagating, scattering and edge mode frequencies, respectively at $35~\si{\kilo\hertz}$, $37~\si{\kilo\hertz}$ and $37.5~ \si{\kilo\hertz}$, shown in Fig.~\ref{fig:HalfSpaceABFlux}. Here we show the displacement of the surface of the half-space; at the positions of the resonators (shown by white points) the displacement is of the attachment point where the resonators are joined to the surface. In Fig.~\ref{fig:HalfSpaceABFlux}(e) we additionally show the vertical displacement ($u_{z}$) of the resonators at a frequency close to that of the edge mode; there is a large amplification (over 100 times) of the vertical displacement of the rods compared to the maximum vertical surface displacement of the half-space, which further corroborates the rods as suitable candidates for harvesters on half-spaces.

\begin{figure}[b!]
    \centering
    \includegraphics[width = 0.9\textwidth]{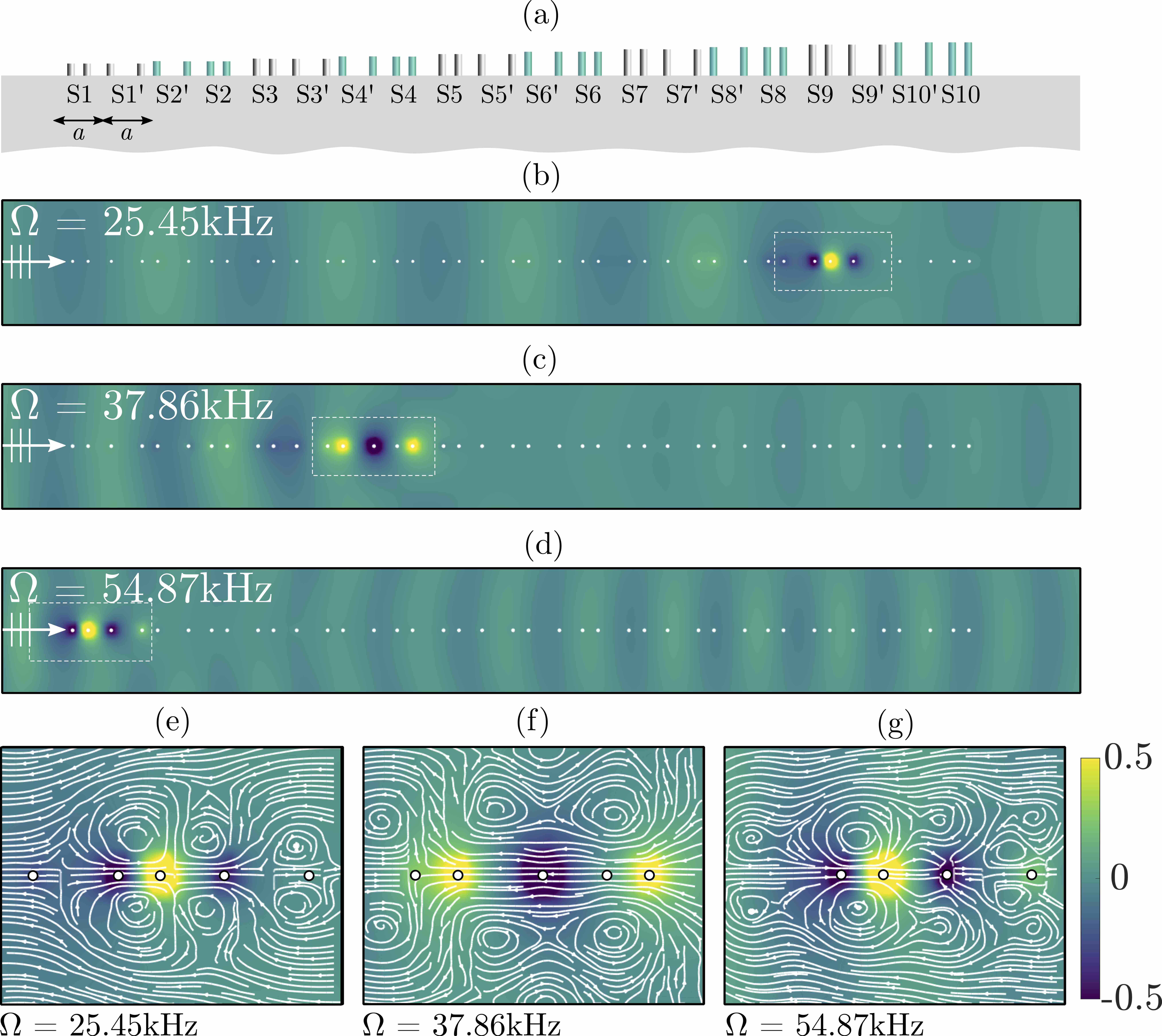}
    \caption{(a) Schematic of Graded SSH metawedge composed of structures in the alternating primed un-primed configuration (shown by grey-green colors of the rods), similar to the beam structure. Rod heights range linearly from $20\si{\milli\meter}$ to $50\si{\milli\meter}$, on an elastic half space of dimensions defined in Fig.~\ref{fig:HSDispersion}. The array is excited, from the left (shown by white arrow), with a broadband Rayleigh wave in the range $10-60~\si{\kilo\hertz}$ with a duration of $38~\si{\milli\second}$, with (b-d) showing a top view of the frequency domain response for three frequencies which correspond to the predicted edge modes obtained from the individual dispersion curves for these heights (as in Fig.~\ref{fig:HSDispersion}), with rod positions marked by white points. (e-g) Show the flux on the surface from the scalar approximation of the surface Rayleigh wave at the corresponding interface positions marked by the dashed white boxes in (b-d). The enhanced amplitude at the predicted interface positions along with the chiral nature of the flux confirms the existence and separate excitation of several edge modes in the graded system.}
    \label{fig:HSFlux}
\end{figure}

To confirm the characteristic chiral flux of the surface edge modes, we calculate the time-averaged flux, only on the surface. To do this, we use a simplified asymptotic approximation for the surface Rayleigh wave, treating the governing equation as a simple scalar wave equation, with the wavespeed corresponding to that of the Rayleigh wave, $c_{r}$. This model has been extensively developed \cite{kaplunov2017asymptotic,wootton2019asymptotic} and adopted in the design of seismic lenses \cite{colombi2016transformation}. It provides a simplification for calculating the flux, $\langle\boldsymbol{F}\rangle$, which now takes the form 
\begin{equation}
    \langle\boldsymbol{F}\rangle = \mathfrak{Im}\left(\phi^*(\nabla\phi)\right),
    \label{eq:scalarflux}
\end{equation}
where $\phi$ is the out of plane displacement of the surface and $^*$ denotes complex conjugation. As such this approximation treats the Rayleigh wave as a scalar surface wave, and allows the nature of the edge modes to be seen, as highlighted in Figs.~\ref{fig:HalfSpaceABFlux},~\ref{fig:HSFlux}.

As for the conventional graded metawedge devices, the design of graded-SSH-metawedge structures for energy harvesting depends on the desired operational frequencies. For conventional metawedges, the height and grading profile is informed by the periodicity and resonances of the individual rods; for efficient harvesting these structures operate by slowing propagating waves to encounter zero group velocity modes at some designed spatial position. Such modes are always present at the band edge by virtue of the Bragg condition. Alternatively, symmetry broken structures can obtain zero group velocity modes within the first BZ \cite{chaplain2020rainbow}. As for graded SSH systems, the edge mode appears at the centre of the band gap \cite{fedorova2019limits}, and as such this allows the tailored design of a stepwise SSH grading to operate over a range of frequencies, as highlighted in Section~\ref{sec:Harvesting}. There is a larger degree of freedom when considering an elastic halfspace compared to, say the KL plate model; surface Rayleigh waves exist independently of any structuring on the array. As such, the broadband excitation of multiple individual edge modes is possible. We highlight this through the design of a graded-SSH-metawedge, shown in Fig.~\ref{fig:HSFlux}(a). The array consists of an Sn-Sn$^\prime$-Sm$^\prime$-Sm configuration as introduced in the case of rods on an elastic beam. The dispersion curves of each individual pair are computed in a similar manner to the example array in Fig.~\ref{fig:HSDispersion}, and the heights selected so that there is an overlap between the longitudinal dispersion curves of the shortest rods with the bandgaps of the tallest rods, which ensures the propagation of the total broadband signal through the array, with lower frequencies travelling furthest through the array. The rod heights range linearly from $20~\si{\milli\meter}$ to $50~\si{\milli\meter}$, all with the same cross sectional thickness defined in Fig.~\ref{fig:HSDispersion}. A broadband Rayleigh wave ($10-60~\si{\kilo\hertz}$) of duration $38 ~\si{\milli\second}$ excites the array, and by frequency domain analysis we show the excitation of several edge modes at their predicted frequencies and interfacial positions between the designed structures (Fig.~\ref{fig:HSFlux}(b-d)). To confirm these are independently excited edge modes we analyse the flux through \eqref{eq:scalarflux}, which shows the chiral nature of this quantity, shown in Fig.~\ref{fig:HSFlux}(e-f).
Despite there only being one cell of the corresponding SSH pairs for each edge mode (i.e. one cell of S1 and one cell of S1'), rather remarkably each predicted edge mode is excited as highlighted by the example in Fig.~\ref{fig:HSFlux}(d) at the extremity of the array. As such, graded-SSH-metawedges on elastic halfspaces have much promise in the design of compact topological rainbow trapping energy harvesters. 
\begin{figure}[h!]
    \centering
    \includegraphics[width = 0.85\textwidth]{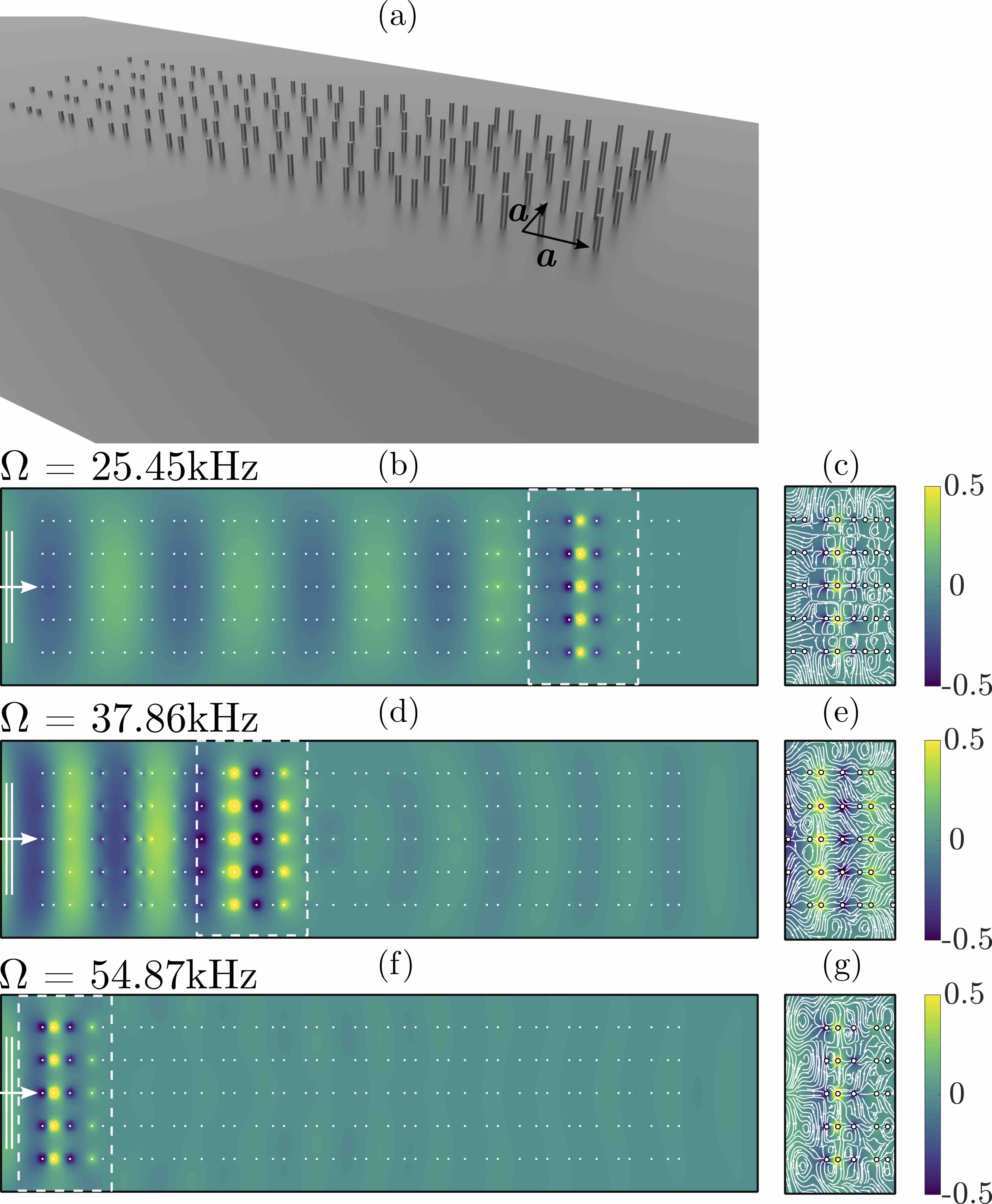}
    \caption{SSH-barrier-metawedge (a) shows a schematic of an SSH-barrier-metawedge, where five rows of the graded array condisered in Fig.\ref{fig:HSFlux} are separated by the array parameter $a$. The barrier array is excited, from the left, with a broadband Rayleigh wave in the range $10-60~\si{\kilo\hertz}$ with a duration of $38~\si{\milli\second}$ (b,d,f) show the multiple edge modes excited in this configuration, matching the positions as in Fig.~\ref{fig:HSFlux}, with their corresponding fluxes shown in (c,e,g). This corroborates the use of the section of the dispersion curves ($\Gamma - X$), which were calculated from a doubly periodic array (Fig.\ref{fig:HSDispersion}) for the prediction of this effect for \textit{normally} incident Rayleigh waves. Again, the edge modes are quickly excited, requiring only one cell of each pair of SSH geometries a a given height, further motivating compact SSH devices.}
    \label{fig:Barrier}
\end{figure}

To further emphasise the utility of these devices, and to better emulate a true metawedge, we extend the device to include several rows of the graded-SSH geometry, representing those considered in \cite{Colombi2017,chaplain2019tailored}. This barrier configuration enables the excitation of more edgemodes in the perpendicular direction. We analyse this system as above, which corroborates with the initial assumptions used to match the edge mode frequency and position from the dispersion curves; the predictions were obtained from the $\Gamma-X$ direction of the two-dimensional dispersion curves and applied to a single, one-dimensional array. In Figure~\ref{fig:Barrier}, the edge modes exist at the same frequencies an spatial positions along the array for normally incident radiation. This arrangement further motivates energy harvesting devices and vibration isolation effects due to the strong confinement of the topological edge modes.

\section{Discussion}

We have successfully implemented the SSH model in a variety of elastic wave regimes, extending the uses of one-dimensional topological modes.
 The coalescence between the, seemingly distinct, SSH topological insulator and the graded metawedge provides a new avenue for topological rainbow devices; these provide significant broadening of the bandwidth over which these one dimensional edge modes can be utilised. Specifically we have shown elastic energy harvesting applications and compared the graded-SSH-metawedge to conventional metawedges, showing a pronounced increase in extracted energy, of approximately $40\%$, due to the strong localisation of these modes. 

Further applications, be it in thin plates, elastic beams or 3D elastic half-spaces, include vibration isolation. These small scale models can be readily scaled up to groundborne and mechanical vibration control  \cite{colombi2016transformation}, and as such we envisage new applications of established topological models by synthesis with other metamaterial structures.

\clearpage

\section*{Acknowledgements}
The authors would like to thank Joseph Sykes, Matthew Proctor and Samuel Palmer for helpful discussions. A.C. was supported by the Ambizione Fellowship PZ00P2-174009. The support of the UK EPSRC through grant EP/T002654/1 is acknowledged as is that of the ERC H2020 FETOpen project BOHEME  under grant agreement No. 863179.


\bibliographystyle{elsarticle-num} 
\bibliography{apssamp.bib}{}


\end{document}